\documentclass[12pt,amsmath,amssymb]{iopart}
\usepackage[dvips]{graphics}
\usepackage{color}

\usepackage{pstricks}

\newcommand{\beq}{\begin{equation}}
\newcommand{\eeq}{\end{equation}}
\newcommand{\bma}{\begin{math}}
\newcommand{\ema}{\end{math}}
\newcommand{\beqa}{\begin{eqnarray}}
\newcommand{\eeqa}{\end{eqnarray}}

\def\opone{\le\textbf{}\textbf{}avevmode\hbox{\small1\kern-3.8pt\normalsize1}}

\newcommand{\be}[1]{     \begin{eqnarray} \mbox{$\label{#1}$}   }

\newcommand{\ee}{\end{eqnarray}}
\newcommand{\pref}[1]{(\ref{#1})}

\newcommand\ket [1] {|#1 \rangle }

\begin{document}

\title{Spin chain description of rotating bosons at $\nu=1$}

\author{Emma Wikberg$^1$, Emil J. Bergholtz$^2$ and Anders Karlhede$^1$}

\address{
$^1$Department of Physics, Stockholm University, \\
AlbaNova University Center, SE-106 91 Stockholm, Sweden\\
$^2$Max Planck Institute  for the Physics of Complex Systems,\\ Noethnitzer Str. 38, 01187 Dresden, Germany}
\eads{\mailto{emma@physto.se}, \mailto{ejb@pks.mpg.de} and
\mailto{ak@physto.se}}

\date{\today}

\begin{abstract}
We consider bosons at Landau level filling $\nu=1$ on a thin torus. In analogy with previous work on fermions at filling $\nu =1/2$, we map the low-energy sector onto a spin-1/2 chain. While the fermionic system may realize the gapless XY-phase, we show that typically this does not happen for the bosonic system. Instead, both delta function and Coulomb interaction lead to gapped phases in the bosonic system, and in particular we identify a phase corresponding to the non-abelian Moore-Read state. In the spin language, the hamiltonian is dominated by a ferromagnetic next-nearest neighbor interaction, which leads to a description consistent with the non-trivial degeneracies of the ground and excited states of this phase of matter. In addition we comment on the similarities and differences of the two systems mentioned above and fermions at $\nu=5/2$.

\end{abstract}

\pacs{73.43.Cd, 71.10.Pm, 75.10.Pq } 

\maketitle

\section{Introduction}\label{introduction}

The equivalence between charged two-dimensional fermions in transverse magnetic fields, and neutral rotating bosons in zero magnetic field has been known for approximately a decade \cite{wilkin,cooper}, see \cite{susanne} for a recent review.  During this time, the connection between fermions in a quantum Hall (QH) system and bosons in very rapidly rotating Bose-Einstein condensates has been a hot subject for theorists and a challenge for experimentalists. For the experimentalists, the main problem lies in being able to rotate the substrate fast enough to get into the QH regime, without making the particles escape the confining potential. Theorists, on the other hand, face basically the same questions as for the ordinary quantum Hall effect---{\it eg} how does one explain why some filling fractions have an energy gap in the spectrum, and how does one understand the nature of the low-energy excitations? In particular, it has been proposed \cite{bosepfaff,boseRR} that these systems may realize non-abelian topological phases \cite{mr,rr,greiter}, thus making them interesting in the context of topological (decoherence free) quantum computation \cite{toprev}.

Theoretically, switching between a fermionic and a bosonic system can be done by means of multiplication of Jastrow factors, $J=\prod_{i<j}(z_{i}-z_{j})$. In the bosonic Laughlin wave function \cite{laughlin}, for example, these factors appear as $\Psi _{\nu =\frac{1}{2m}} \propto J^{2m}$. Multiplying by one extra Jastrow factor changes the filling fraction, $\nu =\frac 1 {2m} \rightarrow \frac 1 {2m+1}$, and, since the Jastrow factor itself is antisymmetric, makes the state fermionic. From this simple analysis one would thus have reason to expect similarities between bosons at $\nu =1$ and fermions at $\nu =1/2$, which, in the lowest Landau level, form a gapless fermi liquid like state \cite{hlr}. However, numerical studies \cite{boseRR,regnault,chang} indicate that the bosonic system on the contrary is gapped for generic forms of the inter-particle interaction, and that it has large overlap with the non-abelian Moore-Read (pfaffian) wave function \cite{mr} (for the fermions this phase is only favorable in a narrow range of interaction space \cite{morf,rh00}, which is however believed to be realized in the second Landau level).

It has recently become clear that it is useful to study the QH problem on a torus, as a function of its circumference, $L_1$. In Landau gauge, this gives a mapping onto a one-dimensional lattice model where the interaction depends on $L_1$. As $L_1\rightarrow 0$ the problem can be exactly diagonalized at any rational filling fraction $\nu=p/q$ \cite{bk2,hierarchy}. The ground states are so called Tao-Thouless states \cite{tt} where the particles have fixed positions as far separated as possible, and the low-lying excitations are fractionally charged domain walls between degenerate ground state configurations. Analytical as well as numerical results support  that these states are  adiabatically connected to the abelian quantum Hall states expected in the bulk ($L_1,L_2\rightarrow \infty$). 

Of course, for some fractions an abelian QH state is not realized in the bulk. In such a case there is a phase transition at finite $L_1$,  to some other phase. In particular, the transition to a gapless state  in the fermionic system at filling $\nu =1/2$  is well understood \cite{bk,bk2}. In analogy to this work, we here define a certain subspace of the bosonic $\nu =1$ many-particle system, in which there is a one-to-one mapping onto a one-dimensional spin-1/2 chain. We then compare the resulting spin hamiltonian with the fermionic ditto in an attempt to obtain a microscopic understanding of the differences between the bosonic and fermionic systems. For realistic interactions, we find that there are qualitative differences between the boson and fermion systems. While the fermions can form a gapless state described by the $XY$-phase in terms of the spin chain, this phase is absent in the bosonic system. Instead, the spin chain description of the boson system leads to a hamiltonian which is dominated by a ferromagnetic next nearest neighbor Ising term. This leads to nontrivial ground state degeneracies and a resulting domain wall description of the quasiparticles which carry the same charge and have the same degeneracies as the non-abelian excitations of the Moore-Read state. We observe a similar, albeit more fragile, phase also for fermions with an interaction appropriate for the $\nu=5/2$ quantum Hall state. An equivalent domain wall description has been found previously \cite{we06,seidel06}. However, in earlier works rather artificial, exactly solvable, model (three-body) interactions have been used to select the ground states. (See also \cite{read06,nonab,eddy} for generalizations to even more exotic states, and \cite{haldanebernevig,wen} for related approaches.) Here, we consider realistic (two-body) interactions and show how the (quasi-) degeneracies spontaneously appear in the system. We achieve this by exact diagonalization studies of small systems, and by interpreting the results in terms of an effective spin chain hamiltonian (which we derive from the microscopic interaction) as outlined above. 

This article is organized as follows. In Section \ref{Thin torus} we introduce a one-dimensional lattice representation of  interacting electrons in a single Landau level. In Section \ref{nuhalf}, we revisit the spin chain description of a half-filled Landau level of fermions, and in Section \ref{bosons} we generalize this to bosons at $\nu=1$, and compare to the fermion case. A brief summary is included in Section \ref{disc}. Definitions and details on the lattice description are given in \ref{model}, and  details on the construction of the spin chain hamiltonian describing the bosonic system are given in \ref{spin}.

\section{Lattice description}\label{Thin torus}

Here we outline how the QH system on a torus can be mapped onto a one-dimensional  lattice model where each site represents a single-particle state with specific momentum. 
We consider a torus with lengths $L_1, L_2$ in the $x-$ and $y-$directions
respectively. Consistent boundary conditions can be enforced when
$L_1L_2=2\pi N_s$  (in units where $\ell=\hbar= c/eB=1$). Here $N_s$ is the number of states in each Landau level, {\it ie} the number of magnetic flux quanta penetrating the surface. In Landau gauge,
$\mathbf{A}=By\mathbf{\hat{x}}$, the one-particle hamiltonian is 
\be{freeham}
H=\frac 1 {2m} ({\bf p} - \frac e c {\bf A})^2=-\frac 1 {2m}[(\partial_x-iy)^2+\partial_y^2] \ ,
\ee
and the states
\begin{equation}
\label{psik}
\psi_{j}=\pi^{-1/4}L_1^{-1/2}\sum_m
e^{i(\frac{2\pi}{L_1}j+mL_2)x}
e^{-(y+\frac{2\pi}{L_1}j+mL_2)^2/2} \ ,
\end{equation}
$j=1,2,...,N_s$, form a basis of one-particle states in the lowest Landau
level. $\psi_j$ is quasiperiodic and
centered along the line $y=-2\pi j/L_1$, thus the $y-$position is given by the $x-$momentum.
This maps the Landau level
onto a one-dimensional lattice model with lattice constant $2\pi/ L_1$.
A basis of many-particle states is given by $|n_1,n_2,\dots, n_{N_s}\rangle$,
where $n_i$ is the number of particles occupying site $i$. For fermions there is either zero or one particle on a specific site ($n_i=0,1$), while several bosons may occupy the same site ($n_i=0,1,2,\ldots$). The filling fraction is defined as $\nu=N/N_s$, where $N=\sum_{i} n_i$ is the total number of particles.

When restricted to a single Landau level,  the hamiltonian consists of the interaction only---there is no kinetic term. Due to momentum conservation, hopping of two particles on the lattice is always symmetrical, {\it ie} the position of the center of mass is preserved. Hence the general two-body hamiltonian takes the form
 \begin{eqnarray} \hat H =\sum_{i=1}^{N_s} \sum_{|m|\leq k \leq N_s/2}V_{km}b^\dagger_{i+m}b^\dagger_{i+k}b_{i+m+k}b_i\equiv \sum_{0\leq m\leq k \leq N_s/2}\hat V_{km}, \label{fullham} \end{eqnarray}
with the (real) matrix elements, $V_{km}$, which depend on the form of the real-space interaction $V(\mathbf r)$. ($H$ is hermitian, which is ensured by $V_{km}=V_{k,-m}$.) $b^\dagger_{i}$ creates a boson or fermion (which one will be clear from the context) in the state $\psi_i$. For more details of this construction, including a definition of the matrix elements $V_{km}$, we refer to \ref{model}.

A crucial observation is that the lattice constant in this model is $2\pi/L_1$, where $L_1$ is the circumference in the $x$-direction of the torus. The extent of a one-particle state in the $y$-direction is of order one, {\it ie} it is independent of $L_1$. Hence, as the torus gets thinner, the overlap between different single-particle states decreases, and the amplitudes of the hopping terms, $V_{km}, m\neq 0$, in the hamiltonian are gradually suppressed. In the limit $L_1\rightarrow 0$, only the repulsive electrostatic terms, $V_{k0}$, remain, and the energy is minimized by keeping the particles as far separated as possible. In the case of $\nu =1/2$ this yields a two-fold degenerate ground state of the form $1010...$, while one gets $1111...$ for $\nu=1$. In general, these thin-limit states are called Tao-Thouless (TT) states \cite{tt} and they always have a gap to the first excited state. In this article we discuss what happens at small, but finite, $L_1$.

\section{Spin chain description of fermions at $\nu =1/2$}\label{nuhalf}

Here we review, and expand, the spin chain description of the half-filled Landau level originally introduced in \cite{bk}.

As already mentioned, in the thin torus limit, $L_1 \rightarrow 0$,  the ground state at $\nu =1/2$ is the TT state $1010...$. Away from this limit the particles will no longer have fixed positions. To describe the physics at small but finite $L_1$ one may define a subspace, $\mathcal H'_f$, of the full fermionic hilbert space $\mathcal H_f$. In $\mathcal H'_f$ there is exactly one particle on each pair of sites $(2i, 2i+1)$ \footnote{The equivalent grouping of sites $(2i-1, 2i)$ gives a copy of the solution presented below.}. Note that the states in $\mathcal H'_f$ have low electrostatic energy by construction. Furthermore, this subspace is naturally mapped  onto a spin-1/2 chain by defining 
\begin{equation} n_{2i}, n_{2i+1}=10\ \leftrightarrow\ \uparrow ,\ \ 
 n_{2i}, n_{2i+1}=01\ \leftrightarrow\ \downarrow. \label{spindef}\end{equation} 

Many of the processes in (\ref{fullham}), including the ones corresponding to the leading hopping term $\hat{V}_{21}$, preserve $\mathcal H'_f$. Truncating the hamiltonian to include only these terms, one finds 
 \beqa \hat H^\prime_f =\sum_{i=1}^{N_s/2}\sum_{k=1}^{N_s/4}\Bigg{[}\frac{\alpha_k}{2} (s^+_is^-_{i+k}+h.c.)+\beta_ks^z_i s^z_{i+k}\Bigg{]},\label{fermham} \eeqa
where
\beqa\alpha_k=2V_{2k,1},\eeqa
\beqa\beta_k=2V_{2k,0}-(1-\delta_{k,N_s/4})V_{2k+1,0}-V_{2k-1,0}.\eeqa
In the restricted hilbert space $\mathcal H'_f$,  the quartic interaction in \pref{fullham} is reduced to the quadratic spin hamiltonian  \pref{fermham}. It is argued in \cite{bk,bk2} that this spin chain hamiltionan describes the system accurately for a range of $L_1$ (for realistic interactions, including Coulomb and short-range interactions, this holds up to $L_1\approx 8$). These arguments are supported by numerical studies, and we refer to the original publications for further details on this.

In Fig. \ref{fermphase} we sketch the phase diagram for the half-filled Landau level as a function of $L_1$.\footnote{The other circumference of the torus can be taken to be infinitely large so that one has a one-dimensional system in the thermodynamic limit.}  As we will explain below, the gapped phase obtained for $L_1<5.3$ (for Coulomb interaction) corresponds to a ferromagnetic Ising phase and the phase just beyond the transition at $L_1\approx 5.3$ is essentially a gapless spin-$\frac 1 2$ XY chain. There is strong numerical evidence that the latter phase is adiabatically connected to the gapless state in the bulk \cite{bk,bk2}.

\begin{figure}[ht]
\begin{center}
\psset{unit=1mm,linewidth=.5mm,dimen=middle}
\begin{pspicture}(0,-1)(10,20)
\psline(-30,-2)(-30,2)
\psline(0,-1.2)(0,1.2)
\psline(-30,0)(60,0)
\psline(58.5,1.5)(60,0)
\psline(58.5,-1.5)(60,0)
\rput(-15,7){$|101010...\rangle$}
\rput(64,-3){$L_1$}
\rput(13,7){XY-phase}
\rput(28,7){$\rightarrow$}
\rput(45,7){Bulk phase}
\rput(0,-5){$5.3$}
\end{pspicture}
\end{center}
\caption{The phase diagram for fermions at $\nu=1/2$, with Coulomb interaction. $L_1$ is one circumference of the torus, the other circumference is infinitely long. }
\label{fermphase}
\end{figure}

\begin{figure}[t]
\begin{center}
\resizebox{!}{110mm}{\includegraphics{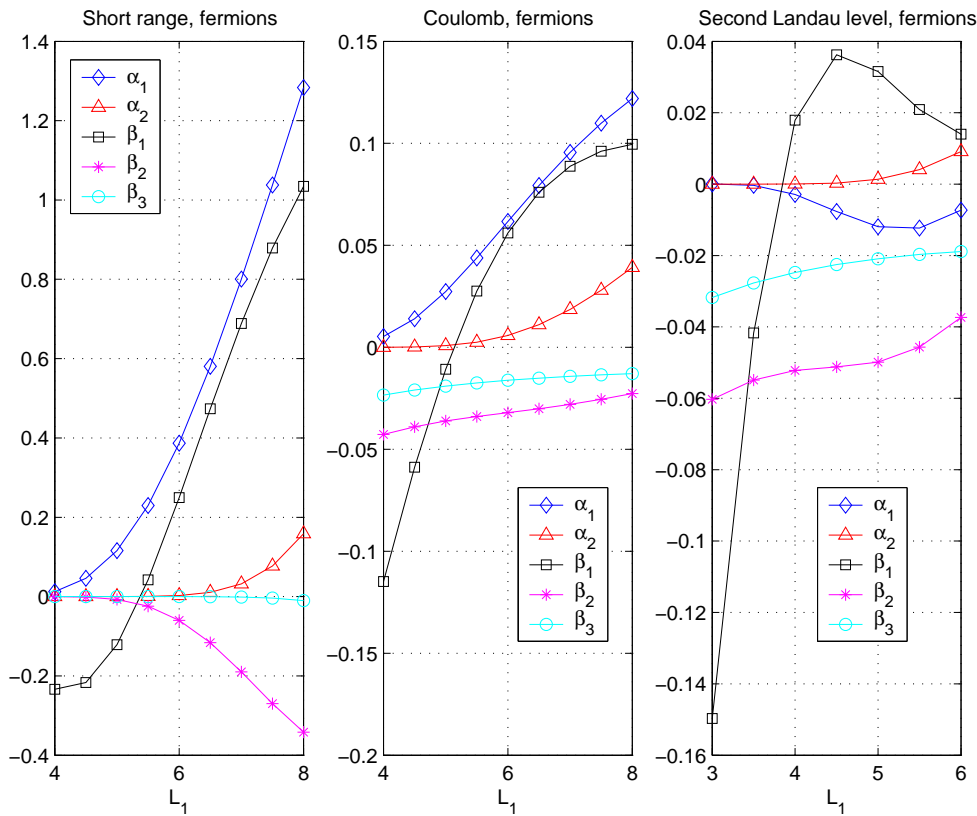}}
\end{center}
\caption{\textit{{\small  The values of the leading coefficients in the spin chain hamiltonian are shown for fermions with different interactions, $V(\mathbf r)$. The leftmost figure corresponds to a short-range interaction, $V(\mathbf r)=\nabla^2\delta_p(\mathbf r)$ and open boundary conditions ({\it ie} a cylinder). In the center we have the Coulomb interaction in the Lowest landau level and to the right the Coulomb interaction in the second Landau level. Here $N_s=16$, but the graphs do not change significantly with particle number.}
}}\label{fig1ferm}
\end{figure}

Let us see how the considerations above are reflected in the functional dependence of the coefficients $\alpha_k$ and $\beta_k$ as we vary $L_1$ in the region of interest ({\it ie} in the region  where numerics tells us that the restriction to $\mathcal H'_f$ is justified). In Fig. \ref{fig1ferm} we show the leading coefficients as  functions of $L_1$ on the thin torus for three different choices of real-space interactions. We see that, when the circumference approaches zero, the hopping terms $\alpha_1$ and $\alpha_2$ tend to zero as expected, and we are left with negative Ising terms, $\beta_k<0$, that favor the TT state $\uparrow\uparrow\uparrow...\leftrightarrow101010...$ \footnote{Of course, the other TT state $\downarrow\downarrow\downarrow...\leftrightarrow010101...$ also has minimal energy in this regime. Note that these states are both included in  $\mathcal H'_f$, and in its translated copy,  $\mathcal H'_{fT}$, hence the total degeneracy in  $\mathcal H_f$ is not $2\times 2=4$, but $2$.}. 
Let us now consider what happens when $L_1$ increases. We see from Fig. \ref{fig1ferm} that one enters a regime where  the ferromagnetic coupling $\beta_1$ that led to the TT ground state weakens and the dominant terms are instead $\beta_2$ and $\alpha_1$---two terms that favor very different ground states. For a short-range interaction, the leftmost panel in Fig. \ref{fig1ferm} shows that the physics is dominated by the $\alpha_1$ term. This is less obvious for the Coulomb interaction in the center panel. However, exact diagonalisation studies  strongly indicate that the system is in the same phase for both these interactions at small but finite $L_1$.\footnote{At  $L_1\approx5.3$ a first order transition from the TT ground state to a state related to the ground state of the $\alpha_1$ term is observed in exact digonalization studies using Coulomb interaction. For a short-range interaction, this transition occurs for slightly smaller $L_1$, and at $L_1\approx 5.3$ the approximation of keeping only $\alpha_1$ is virtually exact, see Fig \ref{fig1ferm} .}  Thus, as a first approximation, we discard all other terms in (\ref{fermham}) and consider:
\begin{equation}
\hat H = \frac{\alpha_{1}} 2\sum_i (s^+_i s^-_{i+1}+h.c.) \ ,
\end{equation}
which is the spin-$\frac 12$ XY chain. This hamiltonian is exactly solvable via a Jordan-Wigner transformation which maps it onto free one-dimensional fermions. These fermions are not the underlying electrons, but rather neutral dipoles---creating one dipole corresponds to creating one electron and annihilating a neighboring electron at the same time ({\it cf} flipping a spin in (\ref{spindef})).  Thus, the quasiparticles are neutral, there is no coupling to the magnetic field and the problem is that of free fermions with a continuous energy spectrum. The ground state is a filled Fermi sea of these fermions and the low-energy excitations are simply particle and hole excitations with respect to this sea.  Of course, all details of the problem are not captured by only taking the $\alpha_1$-term into account. However, as long as the other terms appearing in the hamiltonian are not too big, we stay in the gapless phase and the system is accurately described as a Luttinger liquid. In this way, the mapping onto a spin chain in the regime where the shortest hopping is dominating the hamiltonian,  gives a microscopic insight in why fermions at $\nu =1/2$ form a gapless state with neutral quasiparticles rather than forming a gapped QH system. This formulation is qualitatively in agreement with the standard (mean field) composite fermion \cite{jain} description of this system \cite{hlr}. 

There is also strong numerical evidence that the obtained solution is adiabatically connected to the gapless state in the bulk \cite{bk,bk2}; 
the ground state has a very high overlap with a version of the composite fermion state
\cite{jain} given by Rezayi and Read \cite{rr94}, and this state develops continuously into the
two-dimensional bulk version of the Rezayi-Read state. This establishes the phase diagram displayed in Fig. \ref{fermphase}.

Electrons in the second Landau level (which corresponds to an effectively longer range interaction) have, however, not been studied in this setting before. To this end, we plot the size of the relevant matrix elements for Coulomb interaction in the second Landau level, in the rightmost panel of Fig. \ref{fig1ferm}. We see that $|\beta_2| \gg |\alpha_1|$ in the regime where $\beta_1 \approx 0$, this leads to the physics being different from the lowest Landau level case. We will return to a discussion of this in connection to the boson system below.

\section{Mapping of bosons at $\nu =1$ onto spin chain}\label{bosons}

Inspired by the results obtained above we have performed a similar analysis of bosons at filling $\nu =1$. Also in this case we find a way to map the low-energy sector onto a one-dimensional spin chain, in analogy with the results for fermions at $\nu =1/2$ above. Though, as we shall see, there are two important differences: 1) the restricted Hilbert space is not conserved by {\it any} hopping term, and 2) the hamiltonian is dominated by the next nearest Ising term, $\beta_2$---for a range of $L_1$ and different real-space interactions. This difference sheds light on why $\nu =1$ bosons realize the gapped Moore-Read phase for rather generic interactions \cite{boseRR,regnault,chang}, in contrast to $\nu =1/2$ fermions where this happens only in a small window of interaction space \cite{morf,rh00}, and instead a gapless state forms for sufficiently short range interactions as discussed above.

To achieve a mapping of the torus states onto a spin-1/2 chain, we first restrict to a certain subspace,  $\mathcal H_b'$, within the original Hilbert space $\mathcal H_b$. We define $\mathcal H'_b$ to be the set of states where every $n$ consecutive sites host no more than $n+1$ and no less than $n-1$ particles. Each site then has at most two particles. Furthermore, the restriction excludes two twos, or two zeros, next to each other, or separated by an arbitrarily long string of ones $211...12$ and $011...10$. Now, let every site $n_i$ in such a state split into two new sites $n'_{2i-1}, n'_{2i}$, which share the number of particles of the original site. In other words, let
\[\left\{\begin{array}{lll}n_i=2 & \rightarrow & n'_{2i-1}, n'_{2i}=11\\ 
n_i=0 & \rightarrow & n'_{2i-1}, n'_{2i}=00\\
n_i=1 & \rightarrow & n'_{2i-1}, n'_{2i}=10\ \mathrm{or}\ 01.\end{array}\right.\]
The translation of the 1 follows uniquely from the positions of the zeros and twos in the state. Every 1 to the right of a two or a zero maps as  $1 \rightarrow 01$ and $1\rightarrow 10$ respectively. Whenever a 1 is to the right of a 1, it will be mapped in the same way as the 1 to the left. For completeness, the state with only ones, $111...$, may be defined as  $101010...$. These new lattice states of course have filling fraction one half.

With these definitions, our chosen subspace is identical to the fermionic subspace for $\nu =1/2$ described above, {\it ie} we have states where each pair of sites $(2i,2i+1)$ contains exactly one particle (note that the translated version with sites $(2i-1,2i)$ is not valid here). These can in turn be mapped onto spin-1/2 chains as explained in the previous section; 
\[\left\{\begin{array}{lll} n'_{2i}, n'_{2i+1}=10 & \leftrightarrow & \uparrow\\
n'_{2i}, n'_{2i+1}=01 & \leftrightarrow & \downarrow.\end{array}\right.\]
In other words, $s^z_i=\frac{1}{2}(n'_{2i}-n'_{2i+1})$.
Every site with zero or two particles in the original boson state yields a domain wall between up and down spins in the chain, while the spins corresponding to ones align in the same direction as neighboring spins. For example, 

\begin{equation} 11121110111\rightarrow\downarrow\downarrow\downarrow\uparrow\uparrow\uparrow\uparrow\downarrow\downarrow\downarrow\downarrow , \label{example}  \end{equation}
where we have used the periodic boundary conditions on the torus.

The translation can also be reversed: Starting from an arbitrary spin chain configuration, first
let  
\[\left\{\begin{array}{lll}\uparrow & \rightarrow & n'_{2i}, n'_{2i+1}=10\\
\downarrow & \rightarrow & n'_{2i}, n'_{2i+1}=01\end{array}\right.\]
 or equivalently $n'_{2i}=\frac{1}{2}+s^z_i$, $n'_{2i+1}=\frac{1}{2}-s^z_i$.
Then let
 \[\left\{\begin{array}{lll}n'_{2i-1}, n'_{2i}=11 & \rightarrow & n_i=2 \\ 
n'_{2i-1}, n'_{2i}=00 & \rightarrow & n_i=0 \\
n'_{2i-1}, n'_{2i}=10\ \mathrm{or}\ 01 & \rightarrow & n_i=1.\end{array}\right.\]
to recreate
the bosonic state. The three last equations are equivalent to
\begin{eqnarray}n_i=n'_{2i-1}+n'_{2i}=1+s^z_i-s^z_{i-1},\label{nofsz}\end{eqnarray}
which will be used when we express the bosonic hamiltonian in terms of spin operators.
To conclude, there is a one-to-one mapping between the bosonic subspace $\mathcal H_b'$ and the hilbert space of a spin-1/2 chain (where the two spin-polarized states are defined to be equivalent).

The mapping between the
bosons and the spin chain can be made directly without taking the intermediate step via
fermions. Starting from a subspace boson state, let the spin of a site be
$\downarrow$ ($\uparrow$) if the particle number increases (decreases) to the
right. If the particle number is the same on the site to the right, the spin must be equal to the adjacent spins. This procedure reproduces equation (\ref{example}) above. The inverse map is given by \pref{nofsz},
$n_i=1+s^z_i-s^z_{i-1}$. 

Before we proceed to the effective hamiltonian we discuss the relevance of the subspace $\mathcal H_b'$. We have studied this using exact diagonalization of small systems. As an example, we have diagonalized the hamiltonian \pref{fullham} in the full Hilbert space,  $\mathcal H_b$, on the one hand, and the one restricted to the subspace,  $\mathcal H_b'$, on the other. After diagonalization, the overlap between the respective ground states has been calculated for Coulomb and delta function interaction, and for systems of $N=4,\ 6,\ 8$ particles. In all these cases, for $L_1\approx 4$, the overlap between the two ground states  is above $0.998$. For this $L_1$, the quantum numbers of the ground state has shifted from those of $1111\ldots$ in the thin limit to those of $2020\ldots$, thus the high overlap is a non-trivial result. This situation is reminiscent of the situation for the $\nu=1/2$ fermions. However, there are clear signals that we do not have a phase transition to a gapless state in the bosonic system. First, we observe that the spin polarized state still is very low in energy after the transition (unlike the situation for fermions). Secondly, there are three nearly degenerate states around (and beyond) the transition. Thirdly, each of these three states has the same quantum numbers as one of the Moore-Read states and has a high overlap with this state.  Finally, also in the sector where the two trial states compete, the ground state shows (slightly) higher overlap with the Moore-Read wave function ({\it eg} $0.99$ for $N$=6 and $L_1=4$, Coulomb interaction) than with the Rezayi-Read wave function describing the gapless state ({\it eg} $0.92$ for $N$=6 and $L_1=4$, Coulomb interaction). 

We will now consider the subspace hamiltonian, and investigate whether we can reach an understanding of the bosonic phase diagram by analyzing the spin chain. To proceed we seek a representation in terms of spin operators for all the terms, $\hat{V}_{km}$, that act within $\mathcal H_b'$. This turns out to be more tricky than in the fermionic case studied in the previous section, and the resulting spin hamiltonian contains higher order terms. There are two reasons for this. First,  the bosonic operators are non-local in terms of the (local) spin variables ({\it cf} the inverse of $n_i=1+s^z_i-s^z_{i-1}$), since the mapping of entire domains of ones depend on the particle number to the right (left) of the domain. This implies that a generic (two-body) hopping term, $\hat{V}_{km}$, involves flipping entire domains of spins. Secondly, due to the occupation number dependent action of the bosonic operators ({\it cf}   $b^\dagger_i\ket{\ldots,n_i,\ldots}=\sqrt{n_i+1}\ket{\ldots,n_i+1,\ldots}$ etc) we get more complicated, higher order, terms in the effective hamiltonian. However, these higher order terms have coefficients that are a factor of approximately three smaller than those of the quadratic terms, this makes it reasonable, although not obviously correct, to discard the high-order terms to find a truncated hamiltonian like the one in (\ref{fermham}). The truncated hamiltonian in the bosonic case then becomes
 \beqa \hat H'_b = \sum_{i=1}^{N_s}\sum_{k=1}^{N_s/2}\Big{[}\frac{\alpha_k}{2} (s^+_is^-_{i+k}+h.c.)+\beta_ks^z_i s^z_{i+k}\Big{]} ,\label{spinhambos}\eeqa 
where
\beqa\beta_k=2V_{k0}-(1-\delta_{k,N_s/2}+\delta_{k+1,N_s/2})V_{k+1,0}-(1+\delta_{k,1})V_{k-1,0},\eeqa
\beqa\alpha_1=\frac{1}{2}(4+3\sqrt{2})V_{11}\eeqa
and
\beqa\alpha_k=\frac{1}{8}(17+12\sqrt{2})V_{k1},\ \ k=2, 3, ....\eeqa
Details of the mapping, including the full expressions for all $\hat{V}_{km}$ in terms of spin operators, are given in \ref{spin}. 
 
Let us now consider the hamiltonian in \pref{spinhambos}, and discuss the various phases it possesses on the thin torus to see if we can reach an understanding of why the bosonic system seems to favor the Moore-Read state over the gapless phase  \cite{boseRR,regnault,chang}. For very small $L_1$ the spin chain pictures of the fermion and boson systems are very similar. Here, electrostatic interactions are dominant, leading to an Ising spin hamiltonian with ferromagnetic couplings, $\beta_k<0$. These states are clearly gapped and the elementary excitations are domain walls between spin-polarized domains\footnote{For bosons at $\nu=1$, there is only one ground state, but one may still think of the excitations as domain walls.}. A special case is of course a state where just one spin is flipped relative to the ground state---this is the lowest possible excitation as $L_1\rightarrow 0$.

\begin{figure}[t]
\begin{center}
\resizebox{!}{110mm}{\includegraphics{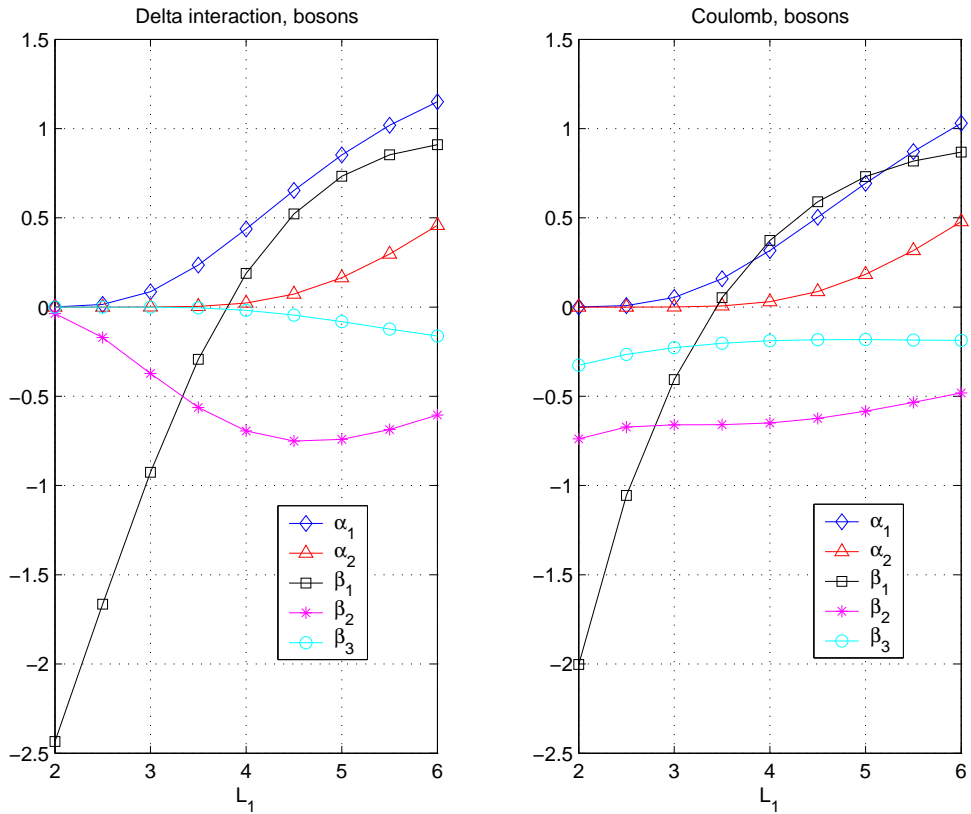}}
\end{center}
\caption{\textit{{\small The values of the leading coefficients in the spin chain hamiltonian are shown for bosons with delta function (left panel) and Coulomb (right panel) interaction. Here $N_s=10$, but the graphs do not change significantly with particle number. }
}}\label{fig2bos}
\end{figure}

\begin{figure}[h!]
\begin{center}
\psset{unit=1mm,linewidth=.5mm,dimen=middle}
\begin{pspicture}(0,-1)(10,20)
\psline(-30,-2)(-30,2)
\psline(-30,0)(60,0)
\psline(58.5,1.5)(60,0)
\psline(58.5,-1.5)(60,0)
\rput(-15,12){$|111111...\rangle$}
\rput(64,-3){$L_1$}
\rput(13,17){$|111111...\rangle$}
\rput(13,12){$|020202...\rangle$}
\rput(13,7){$|202020...\rangle$}
\rput(28,12){$\rightarrow$}
\rput(45,12){Bulk phase}
\rput(-2,-5){$\sim 3.5$}

\psset{unit=1mm,linewidth=.2mm,dimen=middle}
\psline(0,-1.2)(0,1.2)

\end{pspicture}
\end{center}
\caption{Phase diagram for bosons at $\nu=1$. At $L_1 \sim 3.5$ there is a transition from the TT-state 1111..... to a phase initially determined by the $\beta_2$ term. This is then argued to be connected to the Moore-Read state that is believed to describe the bulk phase. (The value  $L_1\sim 3.5$ indicates the approximate value of $L_1$ where the non-trivial (quasi-)degeneracies of the Moore-Read phase appear in the system---the actual level crossing of the ground states appear between 3.7 and 4.0 for both interactions considered here.)
}
\label{bosphase}
\end{figure}
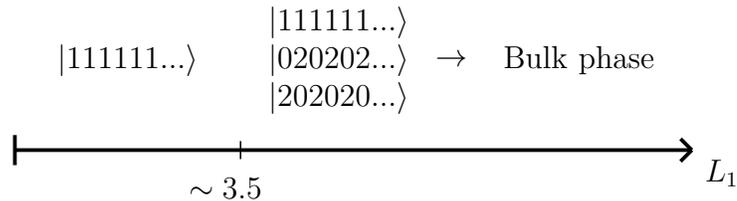

As $L_1$ is increased there is eventually a phase transition. In all cases we have investigated in exact diagonalization, the ground state quantum numbers change from those of the spin-polarized states to a more anti-ferromagnetically looking  state. However, in the bosonic system (in contrast to the situation for fermions) we find that the ferromagnetic state still has low energy after the transition. In fact there are three almost degenerate states, as indicated in Fig. \ref{bosphase}. Moreover, the low lying excitations are essentially built up by finite segments of the different ground states. This is also the situation slightly before the change in ground state quantum numbers, and the transition is thus smoother than in the fermionic case. An example of the structure of the low energy states in this regime is obtained from exact diagonalization for delta interaction at $L_1=3.8$, $N=8$. Here $11111111$ has the lowest energy, which we set to $E_0=0$, and the state 20202020 and its translated version follow with energies $E_1\approx 0.01$ and $E_2\approx 0.04$\footnote{These two energy levels essentially correspond to $2020\ldots \pm 0202\ldots$, small hopping terms break the degeneracy between these states and explains why the levels are not exactly degenerate (the operator that translates all lattice states one site correspond to a good quantum number).}. Then there is a gap  to a number of states with similar energies, all consisting of patterns of domain walls separating the three ground states. For example, we find states of type $11111120$ at $E_3\approx 0.40$, $11202020$ at $E_9\approx 0.49$, and $11110202$ at $E_{11}\approx 0.50$. This structure of the low lying states is qualitatively the same for all $L_1$ around and beyond the level crossing, although the bare Slater determinant states become increasingly dressed with increasing $L_1$. For instance at $L_1=6$ ($L_1=8$) the splitting between the three lowest lying states (they still have the quantum numbers of the Moore-Read ground states) is $0.09$ ($0.05$), while the gap (from the highest of those states) to the domain wall like excitations is $0.35$ ($0.32$). In this context, we also note that on a slightly tilted, or 'rombic', torus (such that it can accommodate a hexagonal unit cell), the quasi degeneracies in the ground state manifold would be promoted to exact ones, see {\it eg} \cite{rh00}.

From Figs. \ref{fig1ferm} and \ref{fig2bos} we get a hint of why the physics is different in the bosonic and fermionic systems as $L_1$ increases. As discovered earlier \cite{bk,bk2}, the ground state of the fermionic system suddenly changes from a gapped TT state to a gapless state for $L_1\sim 5$, for sufficiently short-range interactions (including Coulomb in the lowest Landau level). In the left and center panels of Fig. \ref{fig1ferm}, this is manifested in that $\alpha_1$, {\it ie} the shortest spin-flip term, is the dominating term in the hamiltonian, thus the XY-phase is realized as discussed above. Comparing the boson case,  Fig. \ref{fig2bos}, to the fermion one, Fig. \ref{fig1ferm}, we see that the main difference, in the regime where $\beta_1 \approx 0$, is that $|\beta_2/\alpha_1|$ is substantially larger for the bosons. This is true both for the delta-function and for the Coulomb interaction.

There are two qualitatively different mechanisms responsible for the change in ground state quantum numbers as $L_1$ increases from zero; both the $\alpha_1$ and the $\beta_2$ term are capable of inducing this change, but they lead to drastically different physics. This is corroborated by our exact diagonalization studies as discussed above. 

To understand the physics in the regime where $\beta_1 \approx 0$ and  $|\beta_2/\alpha_1|$ large we now make a (very bold) truncation of the hamiltonian and keep only the $\beta_2$ term. Thus we have 
\begin{equation}\hat H = \beta_{2}\sum_i s^z_i s^z_{i+2},\label{hbos}\end{equation}
which has the three ground states 
\begin{eqnarray} \ket{1}=\uparrow\uparrow\uparrow\uparrow\uparrow\uparrow\uparrow\uparrow\uparrow\uparrow\uparrow\uparrow\cdots\equiv \downarrow\downarrow\downarrow\downarrow\downarrow\downarrow\downarrow\downarrow\downarrow\downarrow\downarrow\cdots
\nonumber\\ \ket{2}=\downarrow\uparrow\downarrow\uparrow\downarrow\uparrow\downarrow\uparrow\downarrow\uparrow\downarrow\uparrow\cdots
\nonumber\\ \ket{\tilde{2}}=\uparrow\downarrow\uparrow\downarrow\uparrow\downarrow\uparrow\downarrow\uparrow\downarrow\uparrow\downarrow\cdots,\label{gs}  \end{eqnarray}
since $\beta_2<0$ in the relevant regime (note that there are only three inequivalent states as the two spin-polarized states are mapped onto the same bosonic state, and are thus equivalent by definition). An excitation with minimal energy can be created by flipping an arbitrary spin in one of the ground states \pref{gs}---this costs an energy $-\beta_2$, and amounts to moving a single particle one site. However, this excitation can be 'fractionalized', at no energy cost,  by replacing the flipped spin by a different ground state; for example,  the states
\begin{eqnarray}\uparrow\uparrow\uparrow\uparrow\uparrow\uparrow\uparrow\uparrow\uparrow\uparrow\uparrow\uparrow\uparrow\uparrow\downarrow\uparrow\uparrow\uparrow\uparrow\uparrow\uparrow\uparrow\uparrow\uparrow\uparrow\uparrow\uparrow\uparrow\uparrow
\nonumber\\ 
\uparrow\uparrow\uparrow\uparrow\uparrow\uparrow\uparrow\uparrow\uparrow\uparrow\uparrow\uparrow\downarrow\uparrow\downarrow\uparrow\downarrow\uparrow\uparrow\uparrow\uparrow\uparrow\uparrow\uparrow\uparrow\uparrow\uparrow\uparrow\uparrow
\nonumber\\ 
\uparrow\uparrow\uparrow\uparrow\uparrow\uparrow\uparrow\uparrow\downarrow\uparrow\downarrow\uparrow\downarrow\uparrow\downarrow\uparrow\downarrow\uparrow\downarrow\uparrow\downarrow\uparrow\uparrow\uparrow\uparrow\uparrow\uparrow\uparrow\uparrow
\label{excitations}  \end{eqnarray}
all have the same (minimal) excitation energy. The two domain walls created this way are quasiparticles with charges $\pm e/2$. The structure of these excitations are in agreement with what we find in our numerical exact diagonalization studies. Note also that the energy gap in the example from the exact diagonalization for $L_1=3.8$ is in agreement with $-\beta_2$ for this value of $L_1$, see Figure \ref{fig2bos}.

Within the spin language, one can only describe quasiparticle-quasihole pairs, as this description completely fixes the filling fraction. However, it is of course possible to go beyond this---the spin language has helped us to understand the elementary excitations of the systems and we can now readily invoke these in a description directly in terms of the original particles. This results in exactly the same domain wall description of the quasiparticles (and holes) as was discovered in \cite{we06,seidel06}.  The present observation that the next-nearest neighbor Ising term ($\beta_2$) naturally appears as a leading term explains the (quasi) degeneracies  for realistic interactions  on the thin torus. Moreover, the fact that the Ising terms dominate in this regime motivates the approximation to keep only the quadratic terms in \pref{spinhambos} as any higher order terms come with coefficients  smaller than the quadratic spin flip terms, $\alpha_k$ (see \ref{spin}). 

Clearly, the truncation of the hamiltonian in \pref{hbos} is very crude. While it certainly is good enough to capture many essential features of the low-energy physics as discussed above, it is not good enough to get a handle on the subtle correlations present in the non-abelian quantum Hall states (at least not without making some extra  assumptions, such as inferring a connection to CFT). Correlation effects may perhaps be unraveled by a more detailed study of the spin chain hamiltonian \pref{spinhambos}, including competing interactions. Spin models with the same symmetries, and including the same terms that appear to be relevant for this problem have been studied earlier \cite{emery}, by means of field theoretical methods and by numerics. However, to the best of our knowledge, no such analysis has yet  been carried out in the parameter regime (large and negative $\beta_2$) encountered here. 
The three ground states of \pref{hbos} suggests that a spin-1 description may also be relevant for the Moore-Read phase. However, we think it is more sensible to start out from a spin-1/2 description, at least in the context of the thin torus, as this allows for an explicit mapping of the microscopic hamiltonian onto a subspace that can be motivated by energetics. On the contrary, it seems hard to obtain a reasonable hamiltonian using a spin-1 mapping. It may still be that a spin-1 picture can shed some light on this problem, and we note that there are similarities with the AKLT spin chain \cite{aklt}. 

It should also be mentioned that the restriction to $\mathcal H_b'$ (or $\mathcal H_f'$) allows for a more accurate description of the antiferromagnetically looking states. For these states the leading quantum fluctuations around the N\'eel states can be described in the restricted hilbert space---this is clearly not the case for the 'ferromagnetic' states. It is not hard to see that the application of {\it any} hopping process, $\hat{V}_{km}$, $m\neq 0$, takes a spin-polarized state to a state outside the spin space. This is something that we see also in our numerical calculations, where the restriction continues to be a very good approximation up to $L_1\approx 6$ (where the overlap between the ground state of the full problem and that of the spin hamiltonian is still as high as 0.968 for a delta function interaction and 0.962 for the Coulomb potential and N=8 bosons) for the 'antiferromagnetic' states while it is a quantitatively reasonable approximation for the ferromagnetic state only in the beginning of the pfaffian phase\footnote{At $L_1=6, N=8$ the leading state configurations, and their weights, are $\ket{\Psi_0}=0.50/0.49(\ket{20202020}+\ket{02020202})+\ldots$, $\ket{\Psi_1}=0.48/0.47(\ket{20202020}-\ket{02020202})+\ldots$ and $\ket{\Psi_2}=0.23/0.23\ket{11111111}+\ldots$ for the delta/Coulomb interactions respectively.}. However, we stress that the simple spin chain picture obtained here is nevertheless relevant for the QH problem; the obtained representation of the low-energy states in terms of domain walls is intimately connected to the conformal field theory description of non-abelian quantum Hall states, and thus encodes the physics of these states assuming the connection to CFT \cite{nonab,wen,eddy}. In this context, we note that the non-abelian statistics has been argued to follow from the domain wall representation by merely assuming adiabatic continuity between the dual $L_1\rightarrow 0$ and $L_2=2\pi N_s/L_1\rightarrow 0$ limits \cite{seidel08}. In the present work we have shown that this domain wall representation is indeed, at least approximately, realized also for realistic two-body interactions (on the thin torus). 
Moreover, the spin chain picture may provide a framework within which one can study correlation effects beyond the overly simplified model in \pref{hbos}. Assuming that we are in the Moore-Read phase, the correlations in all three (or six for the fermions) ground states should have the same nature---thus it is sufficient to be able to understand the correlations in one of the ground states. This may be possible within the spin chain picture, as non-trivial correlations of the antiferromagnetically ordered states are well approximated in spin space also in a region of $L_1$ where the quantum fluctuations are non-negligible.

\section{Conclusion}\label{disc}

We have generalized the mapping of $\nu=1/2$ fermions on a thin torus onto a spin-$1/2$ chain to bosons at $\nu=1$. The resulting spin chain hamiltonians differ---for similar real-space interactions, they lead to qualitatively different physics. For $\nu=1/2$ fermions the hamiltonian is, for sufficiently short-range interactions, dominated by the nearest neighbor spin flip term, leading to a Luttinger liquid ground state, whereas the antiferromagnetic next nearest neighbor Ising term dominates the bosonic case on the thin torus, yielding the known three-fold degenerate Moore-Read state. In a small region in the space of interactions (corresponding to $\nu=5/2$, the second Landau level half filled), this phase is also realized for fermions, where it implies six degenerate ground states. Furthermore, this spin chain description nicely accounts for the emergence of the fractional charge as well as the non-trivial degeneracies of the non-abelian excitations of this phase via the domain wall description. 

It is possible that the full spin chain hamiltonian discussed encodes interesting properties beyond those discussed here. In this context it would be interesting to study the microscopic mechanism driving the gapless state into the Moore-Read phase in more detail.

\ack
We gratefully acknowledge useful correspondence with Susanne Viefers. EJB also acknowledge Mukul Laad for interesting discussions. AK was supported by the Swedish Research Council and by NordForsk.

\appendix
\section{Model}\label{model}

Applying the standard second quantization procedure the interaction becomes
\begin{eqnarray}\hat{H}=\sum_{k_1
 k_2 k_3 k_4}\!\!\!
 V_{k_1 k_2 k_3 k_4}b^{\dagger}_{k_1}b^{\dagger}_{k_2}b_{k_3}b_{k_4} ,\label{hamf}\end{eqnarray}
 where the matrix elements are
\begin{equation}V_{k_1k_2k_3k_4}\!=\frac{1}{2}\int\int \psi_{k_1}^{\dag}(\mathbf{r}_1)\psi_{k_2}^{\dag}(\mathbf{r}_2)V(\mathbf{r}_1-\mathbf{r}_2)\psi_{k_3}(\mathbf{r}_2)\psi_{k_4}(\mathbf{r}_1)  d^{2}r_1d^{2}r_2. \label{matrixelements} \end{equation}
For a periodic interaction, $V(\mathbf{r})$, the matrix elements become 
\begin{eqnarray}
V_{k_1k_2k_3k_4}\!=\!\frac{\delta'_{k_1+k_2,k_3+k_4}}{2N_s}\!\!\!\sum_{q_1,q_2}\delta'_{k_1-k_4,q_1L_1/2\pi}V(\mathbf q)
e^{-\frac{q^2}{2}-i(k_1-k_3)\frac{q_2L_2}{N_s}},\label{matelem}
\end{eqnarray}
where $\delta'$ is the periodic Kronecker delta function (with period $N_s$), $V(\mathbf q)$ is the Fourier transform of $V(\mathbf{r})$
and $q_i=\frac{2\pi n_i}{L_i}$, $n_i=0,\pm 1, \ldots$. For a Coulomb interaction, the 
$\mathbf{q}=0$ term is divergent and must be excluded in (\ref{matelem}); it would be cancelled by
adding a neutralizing background charge.

As a consequence of translation invariance and momentum conservation we can re-write (\ref{hamf}) as
 \begin{eqnarray} \hat H =\sum_{i=1}^{N_s} \sum_{|m|\leq k \leq N_s/2}V_{km}b^\dagger_{i+m}b^\dagger_{i+k}b_{i+m+k}b_i\equiv \sum_{0\leq m\leq k\leq N_s/2}\hat V_{km}, \label{hint} \end{eqnarray}
where
\begin{eqnarray}
V_{km}=\frac 1 {2^{\delta_{k,m}(1+\delta_{k,0})}2^{\delta_{k,N_s/2}}} (V_{n+m,n+k,n+m+k,n}\pm V_{n+m,n+k,n,n+m+k}\nonumber\\+V_{n+k,n+m,n,n+m+k}\pm V_{n+k,n+m,n+m+k,n}) \ .\label{vkm}
\end{eqnarray}
The different signs in \pref{vkm} reflect the statistics of the particles ($+$ for bosons and $-$ for fermions).

The physics of
the interaction can be understood by dividing
$\hat{H}$ into two parts: $V_{k0}$, the electrostatic repulsion
(including exchange)
between two electrons separated $k$ lattice constants, and
$V_{km}$, the amplitude for two particles separated a distance
$k-m$ to hop symmetrically to a separation $k+m$ and vice versa.

\section{Spin chain hamiltonian for the bosons}\label{spin}

Here we provide details on the form of the effective spin chain hamiltonian, $\hat{H}'_b$. First, let us find the spin expressions for all electrostatic and hopping interactions acting within the subspace $\mathcal H_b'$. The electrostatic terms are easiest. Using 
$n_i=1+s^z_i-s^z_{i-1}$ we have (for $k\neq 0$) 
\begin{eqnarray}\hat{V}_{k0}=V_{k0}\sum_ib^\dagger_i b^\dagger_{i-k}b_{i-k}b_{i}=V_{k0}n_in_{i-k}\nonumber\\
=V_{k0}\sum_i(1+s^z_i-s^z_{i-1})(1+s^z_{i-k}-s^z_{i-k-1})
\nonumber\\ =V_{k0}\sum_i(2s^z_is^z_{i+k}-s^z_is^z_{i+k-1}-s^z_is^z_{i+k+1}) + const.,\label{vk0}\end{eqnarray}
and
\begin{eqnarray}\hat{V}_{00}=V_{00}\sum_ib^\dagger_i b^\dagger_{i}b_{i}b_{i}=V_{00}\sum_i n_i(n_i-1)\nonumber\\=V_{00}\sum_i(1+s^z_i-s^z_{i-1})(s^z_i-s^z_{i-1})=-2V_{00}\sum_is^z_is^z_{i+1} + const.\label{v00}.\end{eqnarray}

Deriving the hopping part of the spin hamiltonian is more involved, and we will not show all details on this. It follows that
\begin{eqnarray}b^{\dagger}_{i-k}b^{\dagger}_{i+m}b_{i-k+m}b_i\propto s^+_{i-k}...s^+_{i-k+m-1}s^-_i...s^-_{i+m-1},\end{eqnarray}
where $\propto$ means that the bosonic operators are mapped onto the corresponding spin flips up to occupation number dependent factors. We will now determine these factors. For $k\neq m$ we have
\begin{eqnarray}b^{\dagger}_{i-k}b^{\dagger}_{i+m}b_{i-k+m}b_i|...n_{i-k}...n_{i-k+m}...n_{i}...n_{i+m}..\rangle=\nonumber\\\nonumber\\=\sqrt{n_{i-k}+1}\sqrt{n_{i-k+m}}\sqrt{n_{i}}\sqrt{n_{i+m}+1}\nonumber\\\times|...n_{i-k}+\textrm{\footnotesize{1}}...n_{i-k+m}-\textrm{\footnotesize{1}}...n_{i}-\textrm{\footnotesize{1}}...n_{i+m}+\textrm{\footnotesize{1}}...\rangle.\label{bhop}\end{eqnarray}
The corresponding action in spin space is now easily found. Using $n_i=1+s^z_i-s^z_{i-1}$ \pref{bhop} translates to 
\begin{eqnarray}s^+_{i-k}...s^+_{i-k+m-1}s^-_{i}...s^-_{i+m-1}\sqrt{2+s^z_{i-k}-s^z_{i-k-1}}\nonumber\\\nonumber\\\times\sqrt{1+s^z_{i-k+m}-s^z_{i-k+m-1}}\sqrt{1+s^z_i-s^z_{i-1}}\sqrt{2+s^z_{i+m}-s^z_{i+m-1}}\nonumber\\\nonumber\\\times|...\downarrow_{i-k}...\downarrow_{i-k+m-1}...\uparrow_i...\uparrow_{i+m-1}...\rangle=\nonumber\\\nonumber\\=s^+_{i-k}...s^+_{i-k+m-1}s^-_{i}...s^-_{i+m-1}\nonumber\\\nonumber\\\times\sqrt{\frac{3}{2}-s^z_{i-k-1}}\sqrt{\frac{3}{2}+s^z_{i-k+m}}\sqrt{\frac{3}{2}-s^z_{i-1}}\sqrt{\frac{3}{2}+s^z_{i+m}}\nonumber\\\nonumber\\\times|...\downarrow_{i-k}...\downarrow_{i-k+m-1}...\uparrow_i...\uparrow_{i+m-1}...\rangle, \end{eqnarray}
and identifies the pertinent proportionality factors (which turn out to be spin dependent). 

This result has to be modified a little for $m=k$ and also for $m=k=N_s/2$ because of different bosonic factors in those cases. The general expression for hopping within the subspace becomes
\begin{eqnarray}\hat V _{km} = V_{km}\sum_ib^{\dagger}_{i-k}b^{\dagger}_{i+m}b_{i-k+m}b_i+h.c\nonumber\\=2^{-\delta _{km}(1-\delta_{m,N_s/2})/2} V_{km}\sum_i
s^+_{i-k}... s^+_{i-k+m-1} s^-_{i}... s^-_{i+m-1}\nonumber\\
\times\sqrt{\frac{3}{2} -s^z _{i-k-1}}\sqrt{\frac{3}{2} +s^z _{i-k+m}}\sqrt{\frac{3}{2} -s^z _{i-1}}\sqrt{\frac{3}{2} +s^z _{i+m}} + h.c.,\ m>0.\label{vkm2}\end{eqnarray} 

We see that with all these hopping terms, the hamiltonian is rather complicated. However, we will argue that all terms of order ${\mathcal O} (s^3)$ can be neglected to a first approximation, leading to equation (\ref{spinhambos}).

On the thin torus, the coefficients $V_{km}$ are strongly suppressed with increasing $m$ (the leading behavior is dictated by the overlaps of the single particle states, which  leads to the estimate $V_{km}\sim e^{-2\pi^2m^2/L_1^2}V_{k0}$). Hence, as a first approximation, we let
\begin{eqnarray}\hat H'_b=\sum^{N_s/2}_{k=0}\hat V_{k0}+\sum^{N_s/2}_{k=1}\hat V_{k1}.\end{eqnarray} 

Summing up the electrostatic terms in \pref{vk0} and \pref{v00}, one readily finds 
\beqa \hat H'_{el stat} =\sum_{i=1}^{N_s}\sum_{k=0}^{N_s/2}\hat{V}_{k0} =\sum_{i=1}^{N_s}\sum_{k=1}^{N_s/2}\beta_ks^z_i s^z_{i+k}\label{Helstat}\eeqa 
where $\beta_k=2V_{k0}-(1-\delta_{k,N_s/2}+\delta_{k+1,N_s/2})V_{k+1,0}-(1+\delta_{k,1})V_{k-1,0}$, and the constant term is dropped.

Furthermore, from (\ref{vkm}) we find that the shortest range hopping for bosons,  $\hat V _{11}$, becomes  
\begin{eqnarray}\label{v11}\hat V_{11}=\sqrt 2V_{11}\sum_is^+_{i-1}s^-_i(a^2+ab(s^z_{i+1}-s^z_{i-2}))+{\mathcal O} (s^4)+h.c.,\end{eqnarray}
Where $a=\frac{1+\sqrt{2}}{2}\approx 1.2$ and $b=1-\sqrt 2\approx -0.4$. Also,
\begin{eqnarray}\hat V_{k\neq 1,1}=V_{k1}\sum_is^+_{i-k}s^-_i(a^4+a^3b(s^z_{i-k+1}-s^z_{i-k-1}-s^z_{i-1}+s^z_{i+1}))\nonumber\\+{\mathcal O} (s^4)+h.c..\end{eqnarray}
Here we have used that one may Taylor expand the roots, using $(s^z_i)^2=\frac{1}{4}$, to find
\begin{eqnarray}\sqrt{\frac{3}{2}\pm s^z_i}=a\pm bs^z_i. \label{vk1}\end{eqnarray}
Expanding in the small parameter $|b/a|\approx 1/3$, the leading terms in \pref{v11}, \pref{vk1} are 
 \begin{eqnarray}\hat H'_{hop}=\sum^{N_s/2}_{k=1}\hat V_{k1}=\sum_{i=1}^{N_s}\sum_{k=1}^{N_s/2}\Big{[}\frac{\alpha_k}{2} (s^+_is^-_{i+k}+h.c.)\Big{]} +{\mathcal O} (s^3),\label{Hhop}\end{eqnarray} 
where
\beqa\alpha_1=\frac{1}{2}(4+3\sqrt{2})V_{11}\eeqa
and
\beqa\alpha_k=\frac{1}{8}(17+12\sqrt{2})V_{k1},\ k=2, 3, ....\eeqa
Since $\alpha_k$ in turn are not dominant compared to $\beta_k$ on the thin torus (see Fig. \ref{fig2bos}), we may concentrate on the leading contributions from the hopping terms---at least as long as $\alpha_k$ are not dominating the ${\mathcal O} (s^3)$ terms can be safely ignored. 

Finally, adding (\ref{Helstat}) to (\ref{Hhop}) gives equation (\ref{spinhambos}). The electrostatic terms (\ref{Helstat}) turn out to be the leading terms while the hopping in \pref{Hhop} contain the sub-leading terms on the thin torus. (Of course, if the hopping terms would be dominant the truncation to quadratic terms may not be accurate enough.)

\end{document}